\documentclass[aps,prl,twocolumn,groupedaddress,nofootinbib]{revtex4-1}

\usepackage{graphicx}
\usepackage{mathrsfs}

\usepackage{color}
\begin{document}

\title{Study of the Intermediate Mass Ratio Black Hole Binary Merger up to 1000:1 with Numerical Relativity}


\author{Carlos O. Lousto}
\author{James Healy}

\affiliation{Center for Computational Relativity and Gravitation (CCRG),
School of Mathematical Sciences,
Rochester Institute of Technology, 85 Lomb Memorial Drive, Rochester,
New York 14623}

\date{\today}

\begin{abstract}
We explicitly demonstrate that current numerical relativity techniques
are able to accurately evolve black hole binaries with mass ratios of
the order of 1000:1.  This proof of principle is relevant
for future third generation (3G) gravitational wave detectors and
space mission LISA, as by purely numerical methods we would be able to
accurately compute gravitational waves from the last stages of
black hole mergers, as
directly predicted by general relativity.  We perform a sequence of
simulations in the intermediate to small mass ratio regime,
$m_1^p/m_2^p = 1/7, 1/16, 1/32, 1/64, 1/128, 1/256, 1/512, 1/1024$,
with the small hole starting from rest at a proper distance
$D\approx13M$. We compare these headon full numerical evolutions with
the corresponding semianalytic point particle perturbative results finding an
impressive agreement for the total gravitational radiated energy and
linear momentum as well as for the waveform spectra. We display
numerical convergence of the results and identify the minimal
numerical resolutions required to accurately solve for these very low
amplitude gravitational waves.
This work represents a first step towards the considerable challenge of applying numerical-relativity waveforms to interpreting gravitational-wave observations by LISA and next-generation ground-based gravitational-wave detectors.
\end{abstract}

\maketitle


\section{Introduction\label{sec:introduction}}

There is currently great interest in the binary black hole small mass
ratio regime. On one hand, the direct detection of gravitational waves
from binary black holes since GW150914\cite{LIGOScientific:2016aoc}
and the currently 90 new detections\cite{LIGOScientific:2021djp} by
LIGO-Virgo show consistency with the detailed waveforms predictions of
Numerical Relativity\cite{Campanelli:2005dd,Lovelace:2016uwp} for
comparable mass ratio binaries. Purely full numerical banks of
waveforms~\cite{Healy:2020vre} have been proven to be able and very
effective for estimating binary black hole mergers parameters and
directly applied to the (10+3) gravitational waves signals detected in
the O1/O2 LIGO-Virgo observational runs~\cite{Healy:2020jjs}.

On the other hand, the steady progress towards the establishment of a
LISA launch and operation in the next decade \cite{Gair:2017ynp}, and
the development of third generation (3G) ground detectors
\cite{Purrer:2019jcp,Maggiore:2019uih,Reitze:2019iox}, highlights the
necessity to expand the theoretical studies to much smaller mass
ratios than are currently in use, most of them valid in the comparable
mass-ratio-regime\cite{Healy:2020jjs}.  Scenarios involving
intermediate mass black holes merging with supermassive black holes in
the center of galaxies are targets for LISA
\cite{Amaro-Seoane:2022rxf} and stellar mass black holes merging with
intermediate mass black hole are targets of 3G detectors, also leading
small mass ratios mergers.

While different astrophysical scenarios may consider all sort of
binary black hole systems, for the sake of definiteness, here we will
refer to {\it comparable} mass ratios, $q=m_1/m_2$, as those in the
range $0.1< q\leq1$, {\it intermediate} in the range $0.01< q\leq0.1$,
{\it small} in the range $0.001< q\leq0.01$, and below that in the
{\it near}, $10^{-5}< q\leq10^{-3}$, {\it middle}, $10^{-7}<
q\leq10^{-5}$, and {\it far}, $10^{-9}< q\leq10^{-7}$, {\it extreme} mass
ratio regimes. Most of the numerical simulations focus on the
comparable masses cases, and perturbation theory realm is in that of
the extreme mass ratios. In this work we display techniques that
bridge this gap, performing binary black hole simulations in the
intermediate to small mass ratio binaries regime.

The theoretical interest in the extreme mass ratio (particle
limit) binaries precedes LIGO and LISA conceptions. The pioneering work
of Regge and Wheeler \cite{Regge:1957td} and Zerilli \cite{Zerilli70a}
laid down the formalism for first order perturbations around a single
Schwarzschild black hole, and later work by
Teukolsky\cite{Teukolsky:1973ha} considered the more general Kerr
black hole background.  The consistency of this approach at higher
perturbative orders, including not only radiation reaction but
self-force on the smaller hole (particle), was proven much later by
Mino, Sasaki, and Tanaka\cite{Mino:1996nk} and then by Quinn and
Wald\cite{Quinn:1996am}. This approach proved hard to implement
explicitly, but steady progress has been made since the early headon
collisions\cite{Lousto99b,BL02a} to the current generic orbits (See
Ref.~\cite{Barack:2018yvs} for a current status review).

Numerical Relativity remains the primary method on the forefront of
computing gravitational waves and its results can be used to fit EOB,
Phenom, Surrogate phenomenological models
\cite{Pan:2009wj,Khan:2015jqa,Blackman:2017pcm}.  One of the important
challenges for numerical relativity is the introduction of any new
physical scale to be resolved accurately and, even if mesh refinement
methods are applied, this remains in general computationally
demanding.  A first prototypical study of a 100:1 mass-ratio binary
black hole in \cite{Lousto:2010ut} produced the last two orbits before
merger. The headon collision case was then numerically studied in
\cite{Sperhake:2011ik} (and its $D$-dimensional generalization in
\cite{Cook:2017fec}). In \cite{Husa:2015iqa} it was studied a binary
black hole numerical simulation with spins and a mass ratio 18:1 and
in \cite{Yoo:2022erv} a 15:1 mass ratio case.  A recent revisit to the
small mass ratio problem \cite{Lousto:2020tnb} produced 13 orbits
before merger for a 128:1 mass-ratio binary with moderate computational
resources, leading then to the
possibility of improving effective one body models in the small mass
ratio regime\cite{Nagar:2022icd} and covering the 3G detectors
mass-ratios regime and LISA most conspicuous sources.

Still, LISA could be sensitive to even smaller mass ratio binaries,
two orders of magnitude smaller, deep into the small mass-ratio
regime~\cite{Amaro-Seoane:2022rxf}. This raises the question if
Numerical Relativity can still provide accurate predictions for the
waveforms, trajectories, and final remnant of the merger of such small
mass ratio binary black holes. In this paper we explore the current
limits of the methods of Numerical Relativity, in particular the
moving punctures approach\cite{Campanelli:2005dd}. We will test our
formalism in the regime of up to 1000:1 mass-ratio binaries. For this
first prototypical study we will consider headon collisions from rest and
compare them to the particle limit results (without the need to
include self-force or radiation reaction computations). Our main goal
here is to establish if the moving punctures method can resolve such
extremely small levels of gravitational radiation (even lower
amplitudes than in the orbital case) and if the results show
convergence with increasing numerical resolution towards the
corresponding expected (perturbative) values. This, in turn, can
provide landmarks to reproduce as tests or to use in fits by
semianalytic/phenomenological models~\cite{Nagar:2022icd}.

\section{Numerical Techniques\label{sec:NR}}

In Ref.~\cite{Lousto:2020tnb} we have studied the late inspiral and
merger of small mass ratio binary black holes, reaching a 128:1 case
performing 13 orbits before merger with the use of ${\cal O}(10)$
computational nodes. In order to perform this
simulation we have used the LazEv code\cite{Zlochower:2005bj} with 8th
order spatial finite differences \cite{Lousto:2007rj}, 4th order
Runge-Kutta time integration with a Courant factor $(dt/dx=1/4)$.
Crucially, we used a grid structure developed for the $q=1/15$ simulations in
\cite{Lousto:2018dgd} and adapted for the 128:1 with three additional
refinement levels (15 total) from the boundaries of
the simulation down to the horizon of the smaller hole.

We have performed several convergence and error studies of our
numerical formalism. In Appendix A of Ref.~\cite{Healy:2014yta}, in
Appendix B of Ref.~\cite{Healy:2016lce}, and in
Ref.~\cite{Healy:2017mvh}, we performed convergence studies for
different mass ratios and spins of the binaries.  In in
Ref.~\cite{Lovelace:2016uwp} and Ref.~\cite{Healy:2017abq} we compared
the RIT waveforms with those produced completely independently by the
SXS collaboration finding excellent agreement, convergence towards
each others results and matching of individual modes up to $l=5$.  In
Refs. \cite{Lousto:2020tnb,Rosato:2021jsq} we studied convergence for
an orbiting binary with $q=1/15$ for 10 orbits prior to merger as well
as consistency between radiated and horizon quantities for up to
$q=1/128$ orbiting black hole binaries, and validation versus EOB
models \cite{Nagar:2022icd}. In \cite{Zlochower:2012fk} we have studied
the accuracy of our simulations versus the
Courant factor and concluded that while we can use 1/2 for short runs,
we can ensure long term evolutions with 1/4, and not much gain was
obtained by further reducing it to 1/8. For LIGO-Virgo applications, we
have found that 1/3 was good enough for most of the simulations
\cite{Healy:2017psd,Healy:2019jyf,Healy:2020vre,Healy:2022wdn}.

In the present work we will include several important variations.  We
design a sequence of small mass ratio headon collisions from rest with
decreasing mass ratios by factors of 2 and correspondingly add a new
refinement level of half grid size tight around the small hole, in the
Zeno's approach of \cite{Lousto:2020tnb} and keeping the deepest
refinement level just above the small hole horizon. A second important
difference here is in the use of the numerical gauge. In
Ref.~\cite{Rosato:2021jsq} we have found that the $\eta$ parameter in
the shift $\beta^a$ (Gamma-driver) evolution equation
\begin{equation}\label{eq:gauge}
\partial_t \beta^a = \frac34 \tilde \Gamma^a - \eta(\vec{r},t)\, \beta^a,
\end{equation}
plays an important role in the accuracy of the results and the optimal
use of the grid points to resolve the binary black hole dynamics and
its gravitational radiation. In particular, the use of the $\eta(W)$
driven by the function of the evolved conformal factor $W=e^{-2\Phi}$
in \cite{Lousto:2020tnb} was seen to lead to numerical noise that can
be avoided with the spatial coordinate $\vec{r}$ dependent
$\eta_G(\vec{r})$ choice, while preserving the properties of adapting
to the spacetime around each such dispaired black hole sizes
\begin{eqnarray}\label{eq:etaG}
\eta_G=&&\frac{\mathcal{A}}{m}+\frac {\mathcal{B}}{m_1}\left(\frac{\vec{r}_1(t)^2}{\vec{r}_1(t)^2+\sigma_2^2}\right)^n
e^{-\left|\vec{r}-\vec{r}_1(t)\right|^{2}/\sigma_1^{2}}\nonumber\\
&&+\frac {\mathcal{C}}{m_2}\left(\frac{\vec{r}_2(t)^2}{\vec{r}_2(t)^2+\sigma_1^2}\right)^n
e^{-\left|\vec{r}-\vec{r}_2(t)\right|^{2}/\sigma_2^{2}},
\end{eqnarray}
with ${\cal A}=1$, ${\cal B}=1$, ${\cal C}=1$; $\sigma_1=2\,m_1$,
$\sigma_2=2\,m_2$, $n=2$ used here. $\vec{r}_i(t)$ being the location
of the punctures and $m_i$ being the horizon masses of the holes, with
$m=m_1+m_2$.

In order to test the accuracy of our simulations against the results
of first order perturbation theory, and to study its numerical
convergence, we perform a first prototypical study of the direct
plunge of a small black hole onto a large Schwarzschild black hole
from rest at a reference isotropic coordinate $R_0=10M$, corresponding
to a proper distance from the large hole horizon of
$D/M=\int_{0.5}^{10}(1+1/2x)^2dx=12.9707$ (Here $M$ is the total ADM mass
of the system).

This choice allows us to study the merger of the holes using quadrant
symmetry by placing the small black hole along the z-axis and hence
reduce the numerical grid to one quarter of its full coverage.  The
grid structure of our mesh refinements have a size of the largest box
for all simulations of $\pm400M$.  The number of points between 0 and
400 on the coarsest grid is XXX in nXXX (i.e. n100 has 100 points).
So, the grid spacing on the coarsest is 400/XXX.  The resolution in
the wavezone is $100M/$XXX (i.e. n100 has $M/1.00$, n206 has
$M/2.06$).  The grid around the larger black hole ($m_2$) is fixed at
$\pm1.0M$ in size and is the 9th refinement level.  Therefore the grid
spacing is 400/XXX/$2^8$.  The grid around the small black hole
($m_1$) starts at refinement level 11 for $q=7$ with size
$\pm0.15625M$ and an additional grid is added for each doubling of the
mass ratio with half the size, down to 18 refinement levels for q1024
with size = $\pm0.00125M$.  The minimal grid spacing is then
400/XXX/$2^{(\# \text{refinement levels} - 1)}$.  For q1024 with
resolution of n206, we would have $400/206/2^{17} = 0.000014814$ or a
resolution of $M/67502$.

We are thus able to perform convergence studies at higher global resolutions
than usual with reasonable amounts of computational resources and
running times (for instance, for our resolution, n172, the
q1024 using 18 grid refinement levels simulation took 83 days on 10
dual Intel Xeon 6242 16-core CPUs at 2.8GHz nodes, using a total of
19920 node-hours or 637440 core-hours in our white lagoon CCRG
cluster).

The whole sequence of configurations studied here are
described in Table~\ref{tab:ID} in terms of its initial
parameters. The measured horizon masses follow the expected analytic
Brill-Lindquist~\cite{Brill63,Lousto:1996sx} ratios
$q=m_1/m_2=m_1^p/m_2^p\,(1+m_2^p/2R_0)/(1+m_1^p/2R_0)$, as a function of
the puncture masses $m_i^p$ parameters in the initial data.

\begin{table}
\caption{Initial data parameters for the headon configurations with a
  smaller mass black hole (labeled as 1), and a larger mass spinning
  black hole (labeled as 2). The punctures are located at $\vec r_1 =
  (0,0,z_1)$ and $\vec r_2 = (0,0,z_2)$, have an initial simple proper
  distance\cite{Lousto:2013oza} of $D$, with momenta $P_i=(0,0,0)$ and
  spin $S_i= (0,0,0)$, mass parameters $m_i^p/M$, total ADM mass
  $M_{\rm ADM}=1.0$, the configurations are denoted by qX, where
  X=$m^p_2/m^p_1$, while in the last column $q=m_1 / m_2$ is in
  terms of the horizon masses.}\label{tab:ID}
\begin{ruledtabular}
\begin{tabular}{lcccccc}
Run   & $z_1/M$ & $z_2/M$  & $D/M$ & $m^p_1/M$ & $m^p_2/M$ & $q$ \\
\hline
q7    & -8.7500 & 1.2500 & 13.42 & 0.1250 & 0.8750 & 0.148181  \\
q16   & -9.4118 & 0.5882 & 13.26 & 0.0588 & 0.9412 & 0.065249  \\
q32   & -9.6970 & 0.3030 & 13.18 & 0.0303 & 0.9697 & 0.032716  \\
q64   & -9.8462 & 0.1538 & 13.13 & 0.0154 & 0.9846 & 0.016382  \\
q128  & -9.9225 & 0.0775 & 13.10 & 0.0078 & 0.9922 & 0.008197  \\
q256  & -9.9611 & 0.0389 & 13.09 & 0.0039 & 0.9961 & 0.004100  \\
q512  & -9.9805 & 0.0195 & 13.08 & 0.0019 & 0.9981 & 0.002050  \\
q1024 & -9.9902 & 0.0098 & 13.08 & 0.0010 & 0.9990 & 0.001025  \\
\end{tabular}
\end{ruledtabular}
\end{table}

The extraction of gravitational radiation from the numerical
relativity simulations is performed using the formulas (22) and (23)
from \cite{Campanelli:1998jv} for the energy and linear momentum
radiated, respectively, in terms of the extracted Weyl scalar $\Psi_4$
at the observer location $R_{obs}=113M$.  While in the case of the
particle limit, we evolve the Zerilli equation and the waveform
variable $\psi_{\ell m}$ in the time domain as in \cite{Lousto:1997wf}
and extract the energy and linear momentum as given in formulas (2)
and (4) of Ref. \cite{Lousto:2004pr}. To make a direct comparison of
the numerical and perturbative results we have not removed the
initial spurious radiation from either of the waveforms nor
extrapolated them to infinite observer location.

\section{Simulations' Results\label{sec:results}}

An important goal of this study is to assess the accuracy of our
numerical methods in the so far unexplored regime of a thousand to one
mass ratio black hole binaries. In order to evaluate these estimates
we first perform an internal error analysis by studying the numerical
convergence of gravitational radiation with resolution. Independently,
the second goal is to perform an external comparison of those
radiative quantities with the results of first order perturbations
theory. Once we establish the accuracy of our results we can reliably
start discussing potential correlations between the mode and mass
ratio dependences of the non-linear (numerical relativity) approach to
the linear (perturbative) regime, like our first estimate
$q_{linear}\sim1/(8\ell^2)$ below.

A technical innovation we are going to use here with respect to the
previous simulations \cite{Rosato:2021jsq} is the use of a smoother
gauge as given by Eq.~(\ref{eq:etaG}) that removes the initial
numerical noise.  This is going to play a crucial role here, given the
much lower gravitational wave amplitude levels emitted by the small
mass ratio binaries studied in this paper. In particular,
Eq.~(\ref{eq:etaG}) retains the adaptivity of the gauge to the
different size of the black holes on the grid, but more interestingly
the asymptotic behavior of $\eta_G\to1$ seems to much improve the
extraction of gravitational radiation. 

\begin{figure}
\includegraphics[angle=0,width=\columnwidth]{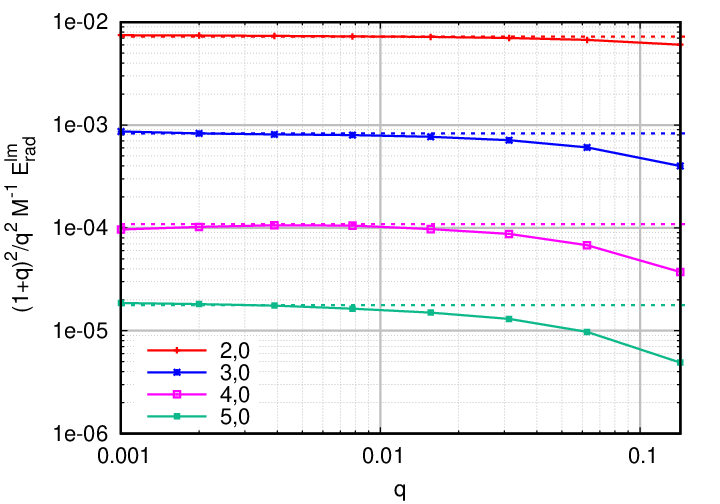}
  \caption{Rescaled radiated energy $E_{\rm rad}/M$, for each mode
    $\ell=2,3,4,5$ for the q7, q16, q32, q64, q128, q256, q512, q1024
    simulations and the particle limit (dotted lines).
  \label{fig:Elmq}}
\end{figure}

Fig.~\ref{fig:Elmq} represents a first display of the results of our
extensive studies based on the highest resolution simulations of
Table~\ref{tab:ID} configurations. In order to perform a direct
comparison we normalize the energy by the leading dependence on the
mass ratio, $m_1^2/M$, thus $(1+q)^2/q^2\,E_{\ell m}/M$, and use their
values at the extraction radius $R_{obs}=113M$.  We first note the
good agreement of the computed (rescaled) radiated energy with the
corresponding particle limit as we approach smaller mass ratios for
all the $\ell$-modes displayed here. We then note that the approach to
the particle limits may depend on the value of $\ell$, roughly in a
sequence $q\sim$ 1/32, 1/64, 1/128, and 1/256 for $\ell=2,3,4,5$
respectively. A critical value of the mass ratio, below which the
particle limit seems to be a very good approximation (within $1\%$) to
the full numerical simulation seems to follow
$q_{linear}\sim1/(8\ell^2)$ to reach the linear regime.

Table~\ref{tab:Evsq} displays our convergence study for the total
radiated energy (summed over $\ell=2$ through $\ell=5$) in the form of
gravitational waves. We use all available runs (except those in
parenthesis) at successive increasing global resolutions $h$ (by
factors of $\approx1.2$) from those labeled n084 to n206 and then use
those values to fit a convergence rate $\gamma$ and its value
extrapolated to infinite resolution of the form
$A_\infty+B\,h^\gamma$.  The results show a high convergence rate, as
expected for the 8th order spatial finite differences
(with q7 being overesolved and n120 underesolving q512 and q1024),
thus approaching the convergence regime, and
extrapolated values close to our highest resolution available serve
a an error measure.
They also display very good agreement with the perturbative results in the
small $q$ cases.
In fact the sequence of the radiated energy
$E_{rad}^{n\infty}$ extrapolated to infinite resolution versus $q$ gives
an approach to the particle limit that can be fitted as an expansion,
$E_{rad}^{fit}=0.00823\,q^2/(1+q)^2-0.0315\,q^3/(1+q)^3+0.0455\,q^4/(1+q)^4$.
From this expression we see that in order to be within $10\%$ the particle
limit we should be about $q<1/32$. 

\begin{table*}
  \caption{The energy radiated, $E_{\rm rad}$, summed over
    $\ell=2,3,4,5$ (and normalized by $M/m_1^2$) for each resolution
    of the qX simulations, starting at D~$\approx13M$ and extracted at
    the radius $R_{obs}=113M$.  All quantities are calculated from the
    gravitational waveforms.  Extrapolation to infinite resolution and
    order of convergence is derived.
\label{tab:Evsq}}
\begin{ruledtabular}
\begin{tabular}{lcccccccl}
Run/resolution   & n084 & n100 & n120 & n144 & n172 & n206 &  n$\infty$ & Order\\
\hline
q7   &0.004885&0.004894&0.004902&0.004905& & &0.004912&2.6\\
q16  &0.006461&0.006513&0.006514&0.006521& & &0.006523&8.3\\
q32  &0.007611&0.007222&0.007330&0.007326& & &0.007272&8.5\\
q64  &(0.008558)&0.008002&0.007688&0.007809& & &0.007885&5.5\\
q128 &(0.008159)&(0.009545)&0.007895&0.008010&0.008044& &0.008059&6.6\\
q256 &(0.011276)&(0.012349)&(0.008778)&0.007943&0.008214&0.008145&0.008123&7.7\\
q512 &(0.037908)&(0.022107)&(0.018858)&0.007693&0.008324&0.008207&0.008182&9.5\\
q1024&(0.099398)&(0.054224)&(0.018103)&0.007582&0.008416&0.008262&0.008229&9.5\\

qparticle &  &  &  &  &  &  &0.008230 &2~\cite{Lousto:1997wf,Lousto:2005ip}\\
\end{tabular}
\end{ruledtabular}
\end{table*}

\begin{figure}
  \includegraphics[angle=0,width=\columnwidth]{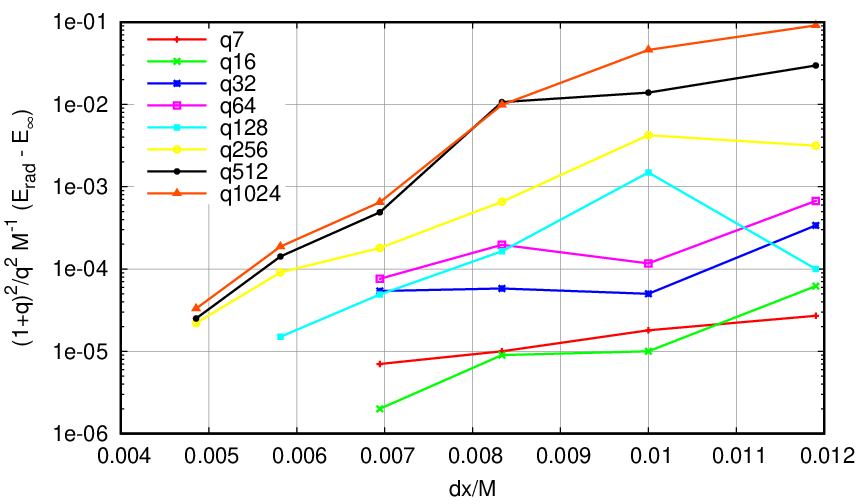}
  \caption{Convergence study of the radiated energy to its extrapolated
to infinite resolution $E_\infty$ values in Table~\ref{tab:Evsq} for the q7, q16, q32, q64, q128, q256, q512, q1024 simulations.
  \label{fig:Einf}}
\end{figure}

A detail of the convergence versus spatial resolution $dx/M$ of the simulation
is displayed in Fig.~\ref{fig:Einf} as the approach of the radiated energy
to its extrapolated to infinite resolution given in Table~\ref{tab:Evsq}.
We have chosen to reach resolutions that at least beat the $10^{-4}$ values of the rescaled quantities, i.e. a $0.1\%$ relative error.

\begin{figure}
\includegraphics[angle=0,width=\columnwidth]{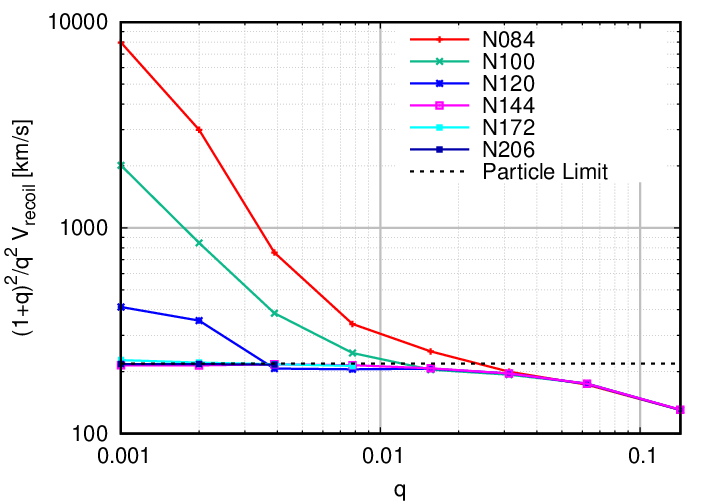}\\
\includegraphics[angle=0,width=\columnwidth]{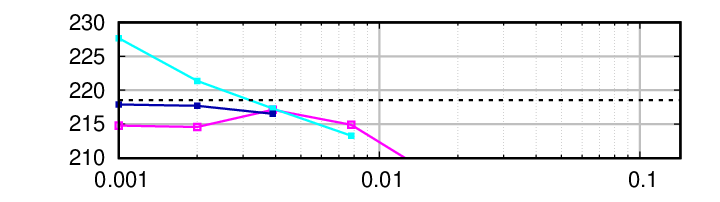}
  \caption{Convergence study of the rescaled recoil velocity,
    $V_{\text{recoil}}$ for the q7, q16, q32, q64, q128, q256, q512,
    q1024 simulations and the particle limit case. With a zoom-in below.
  \label{fig:Vq}}
\end{figure}

In a similar fashion we can study the radiation of linear momentum (in
the headon collision there is no angular momentum to be radiated).  In
this case the coupling of modes $\ell$ and $\ell+1$ in equation (4) of
Ref.~\cite{Lousto:2004pr} makes the computations more sensitive and we
take the opportunity to display in Fig.~\ref{fig:Vq} the dependence of
the results on resolution (in terms of a recoil velocity normalized by
$(1+q)^2/q^2\,V$ and in km/s). This is a more challenging computation
since the recoil comes out as a net difference of linear momentum
radiated rather than the superposition as in the total
energy. Nevertheless the lower panel of Fig.~\ref{fig:Vq} displays a
notable precision with respect to the particle limit and this allows
us, for instance, to assess the minimal resolution required to
reliably compute linear momentum radiated.  This computation of the
recoil velocity thus provides us with a practical assessment of the
minimal global resolution requirement in future computations at very
small binary mass ratios to study the convergence regime.
We will discuss this point further in the next section.

Our results are consistent with those of \cite{Sperhake:2011ik} for the
100:1 mass ratio and extrapolated to infinite observer location. Since,
we extract waveforms at $R_{obs}=113M$ and release the smaller hole from
a finite distance $R_0=10M$ we expect lower values for $E_{rad}$ and $V_r$.

\begin{figure}
\includegraphics[angle=0,width=\columnwidth]{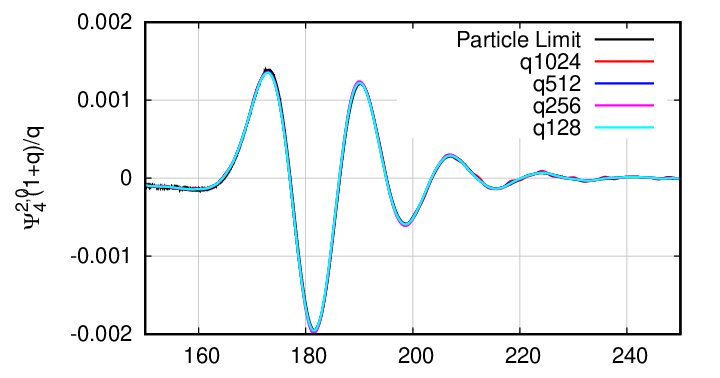}\\
\includegraphics[angle=0,width=\columnwidth]{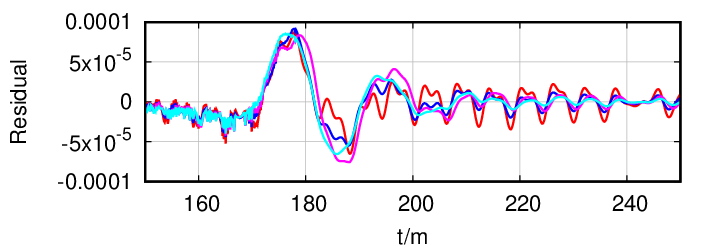}
\caption{Rescaled by $(1+q)/q$ waveforms $\Psi_4$ for the leading
  mode $(\ell,m)=(2,0)$ at the observer location $R_{obs}=113M$
for the q128, q256, q512, and q1024 simulations and the particle limit case.
Bottom panel shows the differences for each mass ratio with the particle limit.
  \label{fig:psi4}}
\end{figure}

A first display of the agreement between the highest available resolution
runs and the particle limit is displayed in Fig.~\ref{fig:psi4}. Those
waveforms are rescaled by the leading dependence with the mass ratio,
$(1+q)/q$, but otherwise not fitted or adjusted. The excellent superposition
of waveforms shows both, the approach to the particle limit above q128,
and the high accuracy of the simulations (note that the rescaling for
q1024 implies a factor of nearly one thousand of amplification).

\begin{figure}
\includegraphics[angle=0,width=\columnwidth]{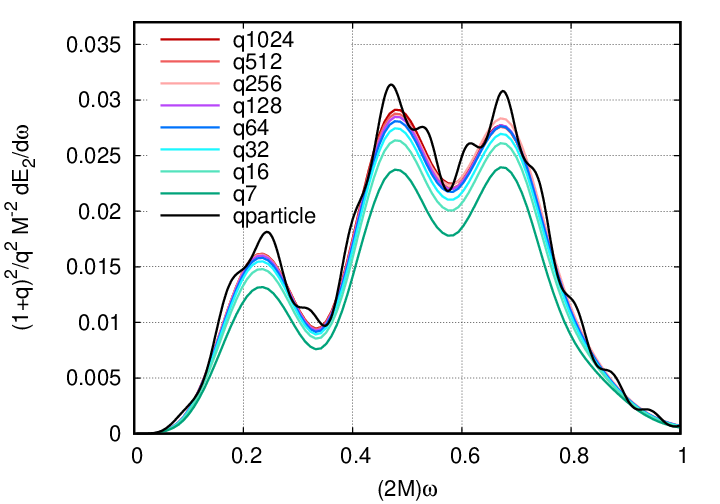}
  \caption{Comparative scaled spectra for for the q7, q16, q32, q64,
    q128, q256, q512, q1024 simulations and the particle limit
    case.
  \label{fig:dEdw}}
\end{figure}

Another detailed feature of the gravitational waves is given by the
spectrum of radiation (for each individual mode, in this case the
leading $(\ell=2,m=0)$, in units of $m_1^2$) as shown in
Fig.~\ref{fig:dEdw}.  This gives us the opportunity to study the
approach to the linear perturbative regime. We see that for q7, q16,
the spectra lies below that of the particle limit (normalized by
$(1+q)^2/q^2/M^2\,dE_{\ell m}/d\omega$) and only at lower values of
$q$ the spectra of the simulations approach that of the particle limit
\cite{Lousto:1996sx}.

Note that the rescaling of $E_{\ell}$ and $V_{\text{recoil}}$ by $\sim
q^{-2}$ implies a precision of one in one million for the q1024 case,
and errors of the order of $1\%$ from the particle limit ensures an
accuracy of the runs of $10^{-8}$ in this low radiation amplitude
headon case.

\section{Conclusions\label{sec:conclusions}}

Our proof of principle in this paper shows that
Numerical Relativity and the moving puncture formalism can effectively
be used to compute small mass ratio binaries (even 1000:1, a so far
unexplored regime, and well into the small mass ratios
territory~\cite{Amaro-Seoane:2022rxf}) and it can accurately describe
those very low amplitude gravitational waveforms.  While the Zeno's
dichotomy approach \cite{Lousto:2020tnb} of adding new mesh refinement
levels to describe the smaller and smaller hole proved to be
appropriate to describe the sources (as monitored by the constancy of
the horizon masses and spins) and the ``conservative'' portion of the
gravitational field in the nonlinear regime, the radiative fields,
that are generated at inter-black holes scales and are extracted in
the asymptotic region at the observer location, scale down with the
mass ratio $q$ and eventually reach such low amplitude values that a
higher global refinement of the grids is required to properly resolve them.
We have thus explored the approach to the convergence regime and have
seen that using the radiated energy as a reference, the simulations
with resolution n084 are sufficient to approximate results for the up
to q16 while the q64 requires at least n100, q256 requires at least
n120, and so on until we reached q1024, requiring at least n144. This
gives a rule-of-thumb for the minimal resolution $h_{min}$ from a base
resolution $h_0=0.84$ to solve accurately a mass ratio binary as small
as $q\approx\frac{1}{16}(h_0/h_{min})^8$.  We confirmed the results with
n172 resolution runs (globally doubling that of n084) as shown
in Table \ref{tab:Evsq}. At further
higher resolutions one may then consider dropping one of the innermost
refinement levels to make simulations more efficient.

We have found that the gauge $\eta_G$ in (\ref{eq:etaG}) notably
improves the numerical noise from our previous $\eta(W)$ used in
\cite{Lousto:2010ut}.  We also found that the $\eta=1$ gauge works
similarly well, suggesting that the lower asymptotic values of $\eta$
allow for a smooth and accurate transport of radiation from the
sources to the extraction regions~\cite{Rosato:2021jsq,Healy:2020iuc},
and that the benefits of an adapted gauge around the horizons is
somewhat taken care of by the additional refinement levels required to
resolve the small hole.  The choice of gauge and the appropriate
minimal global resolutions that warrant the convergence
regime providing valuable quantitative results for the radiation,
orbital trajectories, and remnant of the very small mass ratio binary
black hole mergers, will help us establishing initial parameters
for further orbital evolutions studies.

We have also established for what values of $q$, linear theory begins
to work (or break down) in a merger regime. We have been able to make
a first rough estimate for the threshold value $q_{linear}\leq1/(8\ell^2)$
dependence on the $\ell$-mode.  This deserves further study, including
higher modes and accuracy, in the orbital case. Mode decomposition is
often used in the phenomenological modeling of gravitational waves and
in self-force calculations which include much higher
$\ell$-modes~\cite{Barack:2018yvs}.  Higher $\ell$-modes not only are
smaller in amplitude but of higher frequency content, leading not
only to resolution issues but also to resilience of nonlinearities
(involving derivatives of the fields), as we can infer from our study.

Finally, we note that since the current numerical relativity codes
display a good constant load (weak) scaling, with larger computational
resources they can deal with an increase in resolution of the current
simulations, and perform explorations of the binary's parameter space
(with the advantage of having to deal with increased amplitude of the
gravitational radiation) that may include precession and the spin of
the large hole at essentially no extra delays from the base case that
used ${\cal O}(10)$ computer nodes per run \cite{Lousto:2020tnb}.
Much longer numerical evolutions in time (scaling like ${\cal
  O}(1/q)$), in the intermediate to small mass ratio regime, would
benefit of any future software and hardware speedups by the time of 3G
detectors operation and LISA launch, but our use of a low Courant
factor (1/4) warrants to keep the accuracy of those long term merger
simulations and they could be coupled with a hybridization of the
waveform when the binary is at larger initial separations than those
considered in our pure numerical simulations.  Further work will be
needed in order to determine precisely what advances will be necessary
for binary-black-hole simulations with mass ratios of 1000:1 to
achieve sufficient accuracy (at a feasible cost) for application to
LISA and 3G gravitational-wave detectors.

\begin{acknowledgments}
 The authors gratefully acknowledge the National Science Foundation
(NSF) for financial support from Grants No.\ PHY-1912632 and PHY-2207920.
Computational resources were also provided by the New Horizons, Blue
Sky, Green Prairies, and White Lagoon clusters at the CCRG-Rochester
Institute of Technology, which were supported by NSF grants
No.\ PHY-0722703, No.\ DMS-0820923, No.\ AST-1028087,
No.\ PHY-1229173, No.\ PHY-1726215, and No.\ PHY-2018420.  This work
used the Extreme Science and Engineering Discovery Environment (XSEDE)
[allocation TG-PHY060027N], which is supported by NSF grant
No.\ ACI-1548562 and project PHY20007 Frontera, an NSF-funded
Petascale computing system at the Texas Advanced Computing Center
(TACC).
The authors also thank the referees of this work for numerous useful
suggestions on how to improve the presentation of results.
\end{acknowledgments}

\bibliography{../../../../Bibtex/references}

\begin{thebibliography}{49}%
\makeatletter
\providecommand \@ifxundefined [1]{%
 \@ifx{#1\undefined}
}%
\providecommand \@ifnum [1]{%
 \ifnum #1\expandafter \@firstoftwo
 \else \expandafter \@secondoftwo
 \fi
}%
\providecommand \@ifx [1]{%
 \ifx #1\expandafter \@firstoftwo
 \else \expandafter \@secondoftwo
 \fi
}%
\providecommand \natexlab [1]{#1}%
\providecommand \enquote  [1]{``#1''}%
\providecommand \bibnamefont  [1]{#1}%
\providecommand \bibfnamefont [1]{#1}%
\providecommand \citenamefont [1]{#1}%
\providecommand \href@noop [0]{\@secondoftwo}%
\providecommand \href [0]{\begingroup \@sanitize@url \@href}%
\providecommand \@href[1]{\@@startlink{#1}\@@href}%
\providecommand \@@href[1]{\endgroup#1\@@endlink}%
\providecommand \@sanitize@url [0]{\catcode `\\12\catcode `\$12\catcode
  `\&12\catcode `\#12\catcode `\^12\catcode `\_12\catcode `\%12\relax}%
\providecommand \@@startlink[1]{}%
\providecommand \@@endlink[0]{}%
\providecommand \url  [0]{\begingroup\@sanitize@url \@url }%
\providecommand \@url [1]{\endgroup\@href {#1}{\urlprefix }}%
\providecommand \urlprefix  [0]{URL }%
\providecommand \Eprint [0]{\href }%
\providecommand \doibase [0]{http://dx.doi.org/}%
\providecommand \selectlanguage [0]{\@gobble}%
\providecommand \bibinfo  [0]{\@secondoftwo}%
\providecommand \bibfield  [0]{\@secondoftwo}%
\providecommand \translation [1]{[#1]}%
\providecommand \BibitemOpen [0]{}%
\providecommand \bibitemStop [0]{}%
\providecommand \bibitemNoStop [0]{.\EOS\space}%
\providecommand \EOS [0]{\spacefactor3000\relax}%
\providecommand \BibitemShut  [1]{\csname bibitem#1\endcsname}%
\let\auto@bib@innerbib\@empty
\bibitem [{\citenamefont {Abbott}\ \emph {et~al.}(2016)\citenamefont {Abbott}
  \emph {et~al.}}]{LIGOScientific:2016aoc}%
  \BibitemOpen
  \bibfield  {author} {\bibinfo {author} {\bibfnamefont {B.~P.}\ \bibnamefont
  {Abbott}} \emph {et~al.} (\bibinfo {collaboration} {LIGO Scientific,
  Virgo}),\ }\href {\doibase 10.1103/PhysRevLett.116.061102} {\bibfield
  {journal} {\bibinfo  {journal} {Phys. Rev. Lett.}\ }\textbf {\bibinfo
  {volume} {116}},\ \bibinfo {pages} {061102} (\bibinfo {year} {2016})},\
  \Eprint {http://arxiv.org/abs/1602.03837} {arXiv:1602.03837 [gr-qc]}
  \BibitemShut {NoStop}%
\bibitem [{\citenamefont {Abbott}\ \emph {et~al.}(2021)\citenamefont {Abbott}
  \emph {et~al.}}]{LIGOScientific:2021djp}%
  \BibitemOpen
  \bibfield  {author} {\bibinfo {author} {\bibfnamefont {R.}~\bibnamefont
  {Abbott}} \emph {et~al.} (\bibinfo {collaboration} {LIGO Scientific, VIRGO,
  KAGRA}),\ }\href@noop {} {\  (\bibinfo {year} {2021})},\ \Eprint
  {http://arxiv.org/abs/2111.03606} {arXiv:2111.03606 [gr-qc]} \BibitemShut
  {NoStop}%
\bibitem [{\citenamefont {Campanelli}\ \emph {et~al.}(2006)\citenamefont
  {Campanelli}, \citenamefont {Lousto}, \citenamefont {Marronetti},\ and\
  \citenamefont {Zlochower}}]{Campanelli:2005dd}%
  \BibitemOpen
  \bibfield  {author} {\bibinfo {author} {\bibfnamefont {M.}~\bibnamefont
  {Campanelli}}, \bibinfo {author} {\bibfnamefont {C.~O.}\ \bibnamefont
  {Lousto}}, \bibinfo {author} {\bibfnamefont {P.}~\bibnamefont {Marronetti}},
  \ and\ \bibinfo {author} {\bibfnamefont {Y.}~\bibnamefont {Zlochower}},\
  }\href@noop {} {\bibfield  {journal} {\bibinfo  {journal} {Phys. Rev. Lett.}\
  }\textbf {\bibinfo {volume} {96}},\ \bibinfo {pages} {111101} (\bibinfo
  {year} {2006})},\ \Eprint {http://arxiv.org/abs/gr-qc/0511048}
  {gr-qc/0511048} \BibitemShut {NoStop}%
\bibitem [{\citenamefont {Lovelace}\ \emph {et~al.}(2016)\citenamefont
  {Lovelace} \emph {et~al.}}]{Lovelace:2016uwp}%
  \BibitemOpen
  \bibfield  {author} {\bibinfo {author} {\bibfnamefont {G.}~\bibnamefont
  {Lovelace}} \emph {et~al.},\ }\href {\doibase 10.1088/0264-9381/33/24/244002}
  {\bibfield  {journal} {\bibinfo  {journal} {Class. Quant. Grav.}\ }\textbf
  {\bibinfo {volume} {33}},\ \bibinfo {pages} {244002} (\bibinfo {year}
  {2016})},\ \Eprint {http://arxiv.org/abs/1607.05377} {arXiv:1607.05377
  [gr-qc]} \BibitemShut {NoStop}%
\bibitem [{\citenamefont {Healy}\ and\ \citenamefont
  {Lousto}(2020)}]{Healy:2020vre}%
  \BibitemOpen
  \bibfield  {author} {\bibinfo {author} {\bibfnamefont {J.}~\bibnamefont
  {Healy}}\ and\ \bibinfo {author} {\bibfnamefont {C.~O.}\ \bibnamefont
  {Lousto}},\ }\href {\doibase 10.1103/PhysRevD.102.104018} {\bibfield
  {journal} {\bibinfo  {journal} {Phys. Rev. D}\ }\textbf {\bibinfo {volume}
  {102}},\ \bibinfo {pages} {104018} (\bibinfo {year} {2020})},\ \Eprint
  {http://arxiv.org/abs/2007.07910} {arXiv:2007.07910 [gr-qc]} \BibitemShut
  {NoStop}%
\bibitem [{\citenamefont {Healy}\ \emph
  {et~al.}(2020{\natexlab{a}})\citenamefont {Healy}, \citenamefont {Lousto},
  \citenamefont {Lange},\ and\ \citenamefont {O'Shaughnessy}}]{Healy:2020jjs}%
  \BibitemOpen
  \bibfield  {author} {\bibinfo {author} {\bibfnamefont {J.}~\bibnamefont
  {Healy}}, \bibinfo {author} {\bibfnamefont {C.~O.}\ \bibnamefont {Lousto}},
  \bibinfo {author} {\bibfnamefont {J.}~\bibnamefont {Lange}}, \ and\ \bibinfo
  {author} {\bibfnamefont {R.}~\bibnamefont {O'Shaughnessy}},\ }\href {\doibase
  10.1103/PhysRevD.102.124053} {\bibfield  {journal} {\bibinfo  {journal}
  {Phys. Rev. D}\ }\textbf {\bibinfo {volume} {102}},\ \bibinfo {pages}
  {124053} (\bibinfo {year} {2020}{\natexlab{a}})},\ \Eprint
  {http://arxiv.org/abs/2010.00108} {arXiv:2010.00108 [gr-qc]} \BibitemShut
  {NoStop}%
\bibitem [{\citenamefont {Gair}\ \emph {et~al.}(2017)\citenamefont {Gair},
  \citenamefont {Babak}, \citenamefont {Sesana}, \citenamefont {Amaro-Seoane},
  \citenamefont {Barausse}, \citenamefont {Berry}, \citenamefont {Berti},\ and\
  \citenamefont {Sopuerta}}]{Gair:2017ynp}%
  \BibitemOpen
  \bibfield  {author} {\bibinfo {author} {\bibfnamefont {J.~R.}\ \bibnamefont
  {Gair}}, \bibinfo {author} {\bibfnamefont {S.}~\bibnamefont {Babak}},
  \bibinfo {author} {\bibfnamefont {A.}~\bibnamefont {Sesana}}, \bibinfo
  {author} {\bibfnamefont {P.}~\bibnamefont {Amaro-Seoane}}, \bibinfo {author}
  {\bibfnamefont {E.}~\bibnamefont {Barausse}}, \bibinfo {author}
  {\bibfnamefont {C.~P.}\ \bibnamefont {Berry}}, \bibinfo {author}
  {\bibfnamefont {E.}~\bibnamefont {Berti}}, \ and\ \bibinfo {author}
  {\bibfnamefont {C.}~\bibnamefont {Sopuerta}},\ }\href {\doibase
  10.1088/1742-6596/840/1/012021} {\bibfield  {journal} {\bibinfo  {journal}
  {J. Phys. Conf. Ser.}\ }\textbf {\bibinfo {volume} {840}},\ \bibinfo {pages}
  {012021} (\bibinfo {year} {2017})},\ \Eprint
  {http://arxiv.org/abs/1704.00009} {arXiv:1704.00009 [astro-ph.GA]}
  \BibitemShut {NoStop}%
\bibitem [{\citenamefont {P{\"u}rrer}\ and\ \citenamefont
  {Haster}(2020)}]{Purrer:2019jcp}%
  \BibitemOpen
  \bibfield  {author} {\bibinfo {author} {\bibfnamefont {M.}~\bibnamefont
  {P{\"u}rrer}}\ and\ \bibinfo {author} {\bibfnamefont {C.-J.}\ \bibnamefont
  {Haster}},\ }\href {\doibase 10.1103/PhysRevResearch.2.023151} {\bibfield
  {journal} {\bibinfo  {journal} {Phys. Rev. Res.}\ }\textbf {\bibinfo {volume}
  {2}},\ \bibinfo {pages} {023151} (\bibinfo {year} {2020})},\ \Eprint
  {http://arxiv.org/abs/1912.10055} {arXiv:1912.10055 [gr-qc]} \BibitemShut
  {NoStop}%
\bibitem [{\citenamefont {Maggiore}\ \emph {et~al.}(2020)\citenamefont
  {Maggiore} \emph {et~al.}}]{Maggiore:2019uih}%
  \BibitemOpen
  \bibfield  {author} {\bibinfo {author} {\bibfnamefont {M.}~\bibnamefont
  {Maggiore}} \emph {et~al.},\ }\href {\doibase 10.1088/1475-7516/2020/03/050}
  {\bibfield  {journal} {\bibinfo  {journal} {JCAP}\ }\textbf {\bibinfo
  {volume} {03}},\ \bibinfo {pages} {050} (\bibinfo {year} {2020})},\ \Eprint
  {http://arxiv.org/abs/1912.02622} {arXiv:1912.02622 [astro-ph.CO]}
  \BibitemShut {NoStop}%
\bibitem [{\citenamefont {Reitze}\ \emph {et~al.}(2019)\citenamefont {Reitze}
  \emph {et~al.}}]{Reitze:2019iox}%
  \BibitemOpen
  \bibfield  {author} {\bibinfo {author} {\bibfnamefont {D.}~\bibnamefont
  {Reitze}} \emph {et~al.},\ }\href@noop {} {\bibfield  {journal} {\bibinfo
  {journal} {Bull. Am. Astron. Soc.}\ }\textbf {\bibinfo {volume} {51}},\
  \bibinfo {pages} {035} (\bibinfo {year} {2019})},\ \Eprint
  {http://arxiv.org/abs/1907.04833} {arXiv:1907.04833 [astro-ph.IM]}
  \BibitemShut {NoStop}%
\bibitem [{\citenamefont {Amaro-Seoane}\ \emph {et~al.}(2022)\citenamefont
  {Amaro-Seoane} \emph {et~al.}}]{Amaro-Seoane:2022rxf}%
  \BibitemOpen
  \bibfield  {author} {\bibinfo {author} {\bibfnamefont {P.}~\bibnamefont
  {Amaro-Seoane}} \emph {et~al.},\ }\href@noop {} {\  (\bibinfo {year}
  {2022})},\ \Eprint {http://arxiv.org/abs/2203.06016} {arXiv:2203.06016
  [gr-qc]} \BibitemShut {NoStop}%
\bibitem [{\citenamefont {Regge}\ and\ \citenamefont
  {Wheeler}(1957)}]{Regge:1957td}%
  \BibitemOpen
  \bibfield  {author} {\bibinfo {author} {\bibfnamefont {T.}~\bibnamefont
  {Regge}}\ and\ \bibinfo {author} {\bibfnamefont {J.~A.}\ \bibnamefont
  {Wheeler}},\ }\href {\doibase 10.1103/PhysRev.108.1063} {\bibfield  {journal}
  {\bibinfo  {journal} {Phys. Rev.}\ }\textbf {\bibinfo {volume} {108}},\
  \bibinfo {pages} {1063} (\bibinfo {year} {1957})}\BibitemShut {NoStop}%
\bibitem [{\citenamefont {Zerilli}(1970)}]{Zerilli70a}%
  \BibitemOpen
  \bibfield  {author} {\bibinfo {author} {\bibfnamefont {F.~J.}\ \bibnamefont
  {Zerilli}},\ }\href@noop {} {\bibfield  {journal} {\bibinfo  {journal} {Phys.
  Rev. D.}\ }\textbf {\bibinfo {volume} {2}},\ \bibinfo {pages} {2141}
  (\bibinfo {year} {1970})}\BibitemShut {NoStop}%
\bibitem [{\citenamefont {Teukolsky}(1973)}]{Teukolsky:1973ha}%
  \BibitemOpen
  \bibfield  {author} {\bibinfo {author} {\bibfnamefont {S.~A.}\ \bibnamefont
  {Teukolsky}},\ }\href@noop {} {\bibfield  {journal} {\bibinfo  {journal}
  {Astrophys. J.}\ }\textbf {\bibinfo {volume} {185}},\ \bibinfo {pages} {635}
  (\bibinfo {year} {1973})}\BibitemShut {NoStop}%
\bibitem [{\citenamefont {Mino}\ \emph {et~al.}(1997)\citenamefont {Mino},
  \citenamefont {Sasaki},\ and\ \citenamefont {Tanaka}}]{Mino:1996nk}%
  \BibitemOpen
  \bibfield  {author} {\bibinfo {author} {\bibfnamefont {Y.}~\bibnamefont
  {Mino}}, \bibinfo {author} {\bibfnamefont {M.}~\bibnamefont {Sasaki}}, \ and\
  \bibinfo {author} {\bibfnamefont {T.}~\bibnamefont {Tanaka}},\ }\href
  {\doibase 10.1103/PhysRevD.55.3457} {\bibfield  {journal} {\bibinfo
  {journal} {Phys. Rev.}\ }\textbf {\bibinfo {volume} {D55}},\ \bibinfo {pages}
  {3457} (\bibinfo {year} {1997})},\ \Eprint
  {http://arxiv.org/abs/gr-qc/9606018} {arXiv:gr-qc/9606018} \BibitemShut
  {NoStop}%
\bibitem [{\citenamefont {Quinn}\ and\ \citenamefont
  {Wald}(1997)}]{Quinn:1996am}%
  \BibitemOpen
  \bibfield  {author} {\bibinfo {author} {\bibfnamefont {T.~C.}\ \bibnamefont
  {Quinn}}\ and\ \bibinfo {author} {\bibfnamefont {R.~M.}\ \bibnamefont
  {Wald}},\ }\href {\doibase 10.1103/PhysRevD.56.3381} {\bibfield  {journal}
  {\bibinfo  {journal} {Phys. Rev.}\ }\textbf {\bibinfo {volume} {D56}},\
  \bibinfo {pages} {3381} (\bibinfo {year} {1997})},\ \Eprint
  {http://arxiv.org/abs/gr-qc/9610053} {arXiv:gr-qc/9610053} \BibitemShut
  {NoStop}%
\bibitem [{\citenamefont {Lousto}(2000)}]{Lousto99b}%
  \BibitemOpen
  \bibfield  {author} {\bibinfo {author} {\bibfnamefont {C.~O.}\ \bibnamefont
  {Lousto}},\ }\href@noop {} {\bibfield  {journal} {\bibinfo  {journal} {Phys.
  Rev. Lett.}\ }\textbf {\bibinfo {volume} {84}},\ \bibinfo {pages} {5251}
  (\bibinfo {year} {2000})},\ \Eprint {http://arxiv.org/abs/gr-qc/9912017}
  {gr-qc/9912017} \BibitemShut {NoStop}%
\bibitem [{\citenamefont {Barack}\ and\ \citenamefont {Lousto}(2002)}]{BL02a}%
  \BibitemOpen
  \bibfield  {author} {\bibinfo {author} {\bibfnamefont {L.}~\bibnamefont
  {Barack}}\ and\ \bibinfo {author} {\bibfnamefont {C.~O.}\ \bibnamefont
  {Lousto}},\ }\href@noop {} {\bibfield  {journal} {\bibinfo  {journal} {Phys.
  Rev. D}\ }\textbf {\bibinfo {volume} {66}},\ \bibinfo {pages} {061502}
  (\bibinfo {year} {2002})},\ \Eprint {http://arxiv.org/abs/gr-qc/0205043}
  {gr-qc/0205043} \BibitemShut {NoStop}%
\bibitem [{\citenamefont {Barack}\ and\ \citenamefont
  {Pound}(2019)}]{Barack:2018yvs}%
  \BibitemOpen
  \bibfield  {author} {\bibinfo {author} {\bibfnamefont {L.}~\bibnamefont
  {Barack}}\ and\ \bibinfo {author} {\bibfnamefont {A.}~\bibnamefont {Pound}},\
  }\href {\doibase 10.1088/1361-6633/aae552} {\bibfield  {journal} {\bibinfo
  {journal} {Rept. Prog. Phys.}\ }\textbf {\bibinfo {volume} {82}},\ \bibinfo
  {pages} {016904} (\bibinfo {year} {2019})},\ \Eprint
  {http://arxiv.org/abs/1805.10385} {arXiv:1805.10385 [gr-qc]} \BibitemShut
  {NoStop}%
\bibitem [{\citenamefont {Pan}\ \emph {et~al.}(2010)\citenamefont {Pan} \emph
  {et~al.}}]{Pan:2009wj}%
  \BibitemOpen
  \bibfield  {author} {\bibinfo {author} {\bibfnamefont {Y.}~\bibnamefont
  {Pan}} \emph {et~al.},\ }\href {\doibase 10.1103/PhysRevD.81.084041}
  {\bibfield  {journal} {\bibinfo  {journal} {Phys. Rev.}\ }\textbf {\bibinfo
  {volume} {D81}},\ \bibinfo {pages} {084041} (\bibinfo {year} {2010})},\
  \Eprint {http://arxiv.org/abs/0912.3466} {arXiv:0912.3466 [gr-qc]}
  \BibitemShut {NoStop}%
\bibitem [{\citenamefont {Khan}\ \emph {et~al.}(2016)\citenamefont {Khan},
  \citenamefont {Husa}, \citenamefont {Hannam}, \citenamefont {Ohme},
  \citenamefont {P{\"u}rrer}, \citenamefont {Jim{\'e}nez~Forteza},\ and\
  \citenamefont {Boh{\'e}}}]{Khan:2015jqa}%
  \BibitemOpen
  \bibfield  {author} {\bibinfo {author} {\bibfnamefont {S.}~\bibnamefont
  {Khan}}, \bibinfo {author} {\bibfnamefont {S.}~\bibnamefont {Husa}}, \bibinfo
  {author} {\bibfnamefont {M.}~\bibnamefont {Hannam}}, \bibinfo {author}
  {\bibfnamefont {F.}~\bibnamefont {Ohme}}, \bibinfo {author} {\bibfnamefont
  {M.}~\bibnamefont {P{\"u}rrer}}, \bibinfo {author} {\bibfnamefont
  {X.}~\bibnamefont {Jim{\'e}nez~Forteza}}, \ and\ \bibinfo {author}
  {\bibfnamefont {A.}~\bibnamefont {Boh{\'e}}},\ }\href {\doibase
  10.1103/PhysRevD.93.044007} {\bibfield  {journal} {\bibinfo  {journal} {Phys.
  Rev.}\ }\textbf {\bibinfo {volume} {D93}},\ \bibinfo {pages} {044007}
  (\bibinfo {year} {2016})},\ \Eprint {http://arxiv.org/abs/1508.07253}
  {arXiv:1508.07253 [gr-qc]} \BibitemShut {NoStop}%
\bibitem [{\citenamefont {Blackman}\ \emph {et~al.}(2017)\citenamefont
  {Blackman}, \citenamefont {Field}, \citenamefont {Scheel}, \citenamefont
  {Galley}, \citenamefont {Ott}, \citenamefont {Boyle}, \citenamefont {Kidder},
  \citenamefont {Pfeiffer},\ and\ \citenamefont
  {SzilA!gyi}}]{Blackman:2017pcm}%
  \BibitemOpen
  \bibfield  {author} {\bibinfo {author} {\bibfnamefont {J.}~\bibnamefont
  {Blackman}}, \bibinfo {author} {\bibfnamefont {S.~E.}\ \bibnamefont {Field}},
  \bibinfo {author} {\bibfnamefont {M.~A.}\ \bibnamefont {Scheel}}, \bibinfo
  {author} {\bibfnamefont {C.~R.}\ \bibnamefont {Galley}}, \bibinfo {author}
  {\bibfnamefont {C.~D.}\ \bibnamefont {Ott}}, \bibinfo {author} {\bibfnamefont
  {M.}~\bibnamefont {Boyle}}, \bibinfo {author} {\bibfnamefont {L.~E.}\
  \bibnamefont {Kidder}}, \bibinfo {author} {\bibfnamefont {H.~P.}\
  \bibnamefont {Pfeiffer}}, \ and\ \bibinfo {author} {\bibfnamefont
  {B.}~\bibnamefont {SzilA!gyi}},\ }\href {\doibase 10.1103/PhysRevD.96.024058}
  {\bibfield  {journal} {\bibinfo  {journal} {Phys. Rev.}\ }\textbf {\bibinfo
  {volume} {D96}},\ \bibinfo {pages} {024058} (\bibinfo {year} {2017})},\
  \Eprint {http://arxiv.org/abs/1705.07089} {arXiv:1705.07089 [gr-qc]}
  \BibitemShut {NoStop}%
\bibitem [{\citenamefont {Lousto}\ and\ \citenamefont
  {Zlochower}(2011)}]{Lousto:2010ut}%
  \BibitemOpen
  \bibfield  {author} {\bibinfo {author} {\bibfnamefont {C.~O.}\ \bibnamefont
  {Lousto}}\ and\ \bibinfo {author} {\bibfnamefont {Y.}~\bibnamefont
  {Zlochower}},\ }\href {\doibase 10.1103/PhysRevLett.106.041101} {\bibfield
  {journal} {\bibinfo  {journal} {Phys. Rev. Lett.}\ }\textbf {\bibinfo
  {volume} {106}},\ \bibinfo {pages} {041101} (\bibinfo {year} {2011})},\
  \Eprint {http://arxiv.org/abs/1009.0292} {arXiv:1009.0292 [gr-qc]}
  \BibitemShut {NoStop}%
\bibitem [{\citenamefont {Sperhake}\ \emph {et~al.}(2011)\citenamefont
  {Sperhake}, \citenamefont {Cardoso}, \citenamefont {Ott}, \citenamefont
  {Schnetter},\ and\ \citenamefont {Witek}}]{Sperhake:2011ik}%
  \BibitemOpen
  \bibfield  {author} {\bibinfo {author} {\bibfnamefont {U.}~\bibnamefont
  {Sperhake}}, \bibinfo {author} {\bibfnamefont {V.}~\bibnamefont {Cardoso}},
  \bibinfo {author} {\bibfnamefont {C.~D.}\ \bibnamefont {Ott}}, \bibinfo
  {author} {\bibfnamefont {E.}~\bibnamefont {Schnetter}}, \ and\ \bibinfo
  {author} {\bibfnamefont {H.}~\bibnamefont {Witek}},\ }\href {\doibase
  10.1103/PhysRevD.84.084038} {\bibfield  {journal} {\bibinfo  {journal} {Phys.
  Rev.}\ }\textbf {\bibinfo {volume} {D84}},\ \bibinfo {pages} {084038}
  (\bibinfo {year} {2011})},\ \Eprint {http://arxiv.org/abs/1105.5391}
  {arXiv:1105.5391 [gr-qc]} \BibitemShut {NoStop}%
\bibitem [{\citenamefont {Cook}\ \emph {et~al.}(2017)\citenamefont {Cook},
  \citenamefont {Sperhake}, \citenamefont {Berti},\ and\ \citenamefont
  {Cardoso}}]{Cook:2017fec}%
  \BibitemOpen
  \bibfield  {author} {\bibinfo {author} {\bibfnamefont {W.~G.}\ \bibnamefont
  {Cook}}, \bibinfo {author} {\bibfnamefont {U.}~\bibnamefont {Sperhake}},
  \bibinfo {author} {\bibfnamefont {E.}~\bibnamefont {Berti}}, \ and\ \bibinfo
  {author} {\bibfnamefont {V.}~\bibnamefont {Cardoso}},\ }\href {\doibase
  10.1103/PhysRevD.96.124006} {\bibfield  {journal} {\bibinfo  {journal} {Phys.
  Rev. D}\ }\textbf {\bibinfo {volume} {96}},\ \bibinfo {pages} {124006}
  (\bibinfo {year} {2017})},\ \Eprint {http://arxiv.org/abs/1709.10514}
  {arXiv:1709.10514 [gr-qc]} \BibitemShut {NoStop}%
\bibitem [{\citenamefont {Husa}\ \emph {et~al.}(2016)\citenamefont {Husa},
  \citenamefont {Khan}, \citenamefont {Hannam}, \citenamefont {P{\"u}rrer},
  \citenamefont {Ohme}, \citenamefont {Jim{\'e}nez~Forteza},\ and\
  \citenamefont {Boh{\'e}}}]{Husa:2015iqa}%
  \BibitemOpen
  \bibfield  {author} {\bibinfo {author} {\bibfnamefont {S.}~\bibnamefont
  {Husa}}, \bibinfo {author} {\bibfnamefont {S.}~\bibnamefont {Khan}}, \bibinfo
  {author} {\bibfnamefont {M.}~\bibnamefont {Hannam}}, \bibinfo {author}
  {\bibfnamefont {M.}~\bibnamefont {P{\"u}rrer}}, \bibinfo {author}
  {\bibfnamefont {F.}~\bibnamefont {Ohme}}, \bibinfo {author} {\bibfnamefont
  {X.}~\bibnamefont {Jim{\'e}nez~Forteza}}, \ and\ \bibinfo {author}
  {\bibfnamefont {A.}~\bibnamefont {Boh{\'e}}},\ }\href {\doibase
  10.1103/PhysRevD.93.044006} {\bibfield  {journal} {\bibinfo  {journal} {Phys.
  Rev.}\ }\textbf {\bibinfo {volume} {D93}},\ \bibinfo {pages} {044006}
  (\bibinfo {year} {2016})},\ \Eprint {http://arxiv.org/abs/1508.07250}
  {arXiv:1508.07250 [gr-qc]} \BibitemShut {NoStop}%
\bibitem [{\citenamefont {Yoo}\ \emph {et~al.}(2022)\citenamefont {Yoo},
  \citenamefont {Varma}, \citenamefont {Giesler}, \citenamefont {Scheel},
  \citenamefont {Haster}, \citenamefont {Pfeiffer}, \citenamefont {Kidder},\
  and\ \citenamefont {Boyle}}]{Yoo:2022erv}%
  \BibitemOpen
  \bibfield  {author} {\bibinfo {author} {\bibfnamefont {J.}~\bibnamefont
  {Yoo}}, \bibinfo {author} {\bibfnamefont {V.}~\bibnamefont {Varma}}, \bibinfo
  {author} {\bibfnamefont {M.}~\bibnamefont {Giesler}}, \bibinfo {author}
  {\bibfnamefont {M.~A.}\ \bibnamefont {Scheel}}, \bibinfo {author}
  {\bibfnamefont {C.-J.}\ \bibnamefont {Haster}}, \bibinfo {author}
  {\bibfnamefont {H.~P.}\ \bibnamefont {Pfeiffer}}, \bibinfo {author}
  {\bibfnamefont {L.~E.}\ \bibnamefont {Kidder}}, \ and\ \bibinfo {author}
  {\bibfnamefont {M.}~\bibnamefont {Boyle}},\ }\href {\doibase
  10.1103/PhysRevD.106.044001} {\bibfield  {journal} {\bibinfo  {journal}
  {Phys. Rev. D}\ }\textbf {\bibinfo {volume} {106}},\ \bibinfo {pages}
  {044001} (\bibinfo {year} {2022})},\ \Eprint
  {http://arxiv.org/abs/2203.10109} {arXiv:2203.10109 [gr-qc]} \BibitemShut
  {NoStop}%
\bibitem [{\citenamefont {Lousto}\ and\ \citenamefont
  {Healy}(2020)}]{Lousto:2020tnb}%
  \BibitemOpen
  \bibfield  {author} {\bibinfo {author} {\bibfnamefont {C.~O.}\ \bibnamefont
  {Lousto}}\ and\ \bibinfo {author} {\bibfnamefont {J.}~\bibnamefont {Healy}},\
  }\href {\doibase 10.1103/PhysRevLett.125.191102} {\bibfield  {journal}
  {\bibinfo  {journal} {Phys. Rev. Lett.}\ }\textbf {\bibinfo {volume} {125}},\
  \bibinfo {pages} {191102} (\bibinfo {year} {2020})},\ \Eprint
  {http://arxiv.org/abs/2006.04818} {arXiv:2006.04818 [gr-qc]} \BibitemShut
  {NoStop}%
\bibitem [{\citenamefont {Nagar}\ \emph {et~al.}(2022)\citenamefont {Nagar},
  \citenamefont {Healy}, \citenamefont {Lousto}, \citenamefont {Bernuzzi},\
  and\ \citenamefont {Albertini}}]{Nagar:2022icd}%
  \BibitemOpen
  \bibfield  {author} {\bibinfo {author} {\bibfnamefont {A.}~\bibnamefont
  {Nagar}}, \bibinfo {author} {\bibfnamefont {J.}~\bibnamefont {Healy}},
  \bibinfo {author} {\bibfnamefont {C.~O.}\ \bibnamefont {Lousto}}, \bibinfo
  {author} {\bibfnamefont {S.}~\bibnamefont {Bernuzzi}}, \ and\ \bibinfo
  {author} {\bibfnamefont {A.}~\bibnamefont {Albertini}},\ }\href@noop {} {\
  (\bibinfo {year} {2022})},\ \Eprint {http://arxiv.org/abs/2202.05643}
  {arXiv:2202.05643 [gr-qc]} \BibitemShut {NoStop}%
\bibitem [{\citenamefont {Zlochower}\ \emph {et~al.}(2005)\citenamefont
  {Zlochower}, \citenamefont {Baker}, \citenamefont {Campanelli},\ and\
  \citenamefont {Lousto}}]{Zlochower:2005bj}%
  \BibitemOpen
  \bibfield  {author} {\bibinfo {author} {\bibfnamefont {Y.}~\bibnamefont
  {Zlochower}}, \bibinfo {author} {\bibfnamefont {J.~G.}\ \bibnamefont
  {Baker}}, \bibinfo {author} {\bibfnamefont {M.}~\bibnamefont {Campanelli}}, \
  and\ \bibinfo {author} {\bibfnamefont {C.~O.}\ \bibnamefont {Lousto}},\
  }\href {\doibase 10.1103/PhysRevD.72.024021} {\bibfield  {journal} {\bibinfo
  {journal} {Phys. Rev.}\ }\textbf {\bibinfo {volume} {D72}},\ \bibinfo {pages}
  {024021} (\bibinfo {year} {2005})},\ \Eprint
  {http://arxiv.org/abs/gr-qc/0505055} {arXiv:gr-qc/0505055} \BibitemShut
  {NoStop}%
\bibitem [{\citenamefont {Lousto}\ and\ \citenamefont
  {Zlochower}(2008)}]{Lousto:2007rj}%
  \BibitemOpen
  \bibfield  {author} {\bibinfo {author} {\bibfnamefont {C.~O.}\ \bibnamefont
  {Lousto}}\ and\ \bibinfo {author} {\bibfnamefont {Y.}~\bibnamefont
  {Zlochower}},\ }\href {\doibase 10.1103/PhysRevD.77.024034} {\bibfield
  {journal} {\bibinfo  {journal} {Phys. Rev.}\ }\textbf {\bibinfo {volume}
  {D77}},\ \bibinfo {pages} {024034} (\bibinfo {year} {2008})},\ \Eprint
  {http://arxiv.org/abs/0711.1165} {arXiv:0711.1165 [gr-qc]} \BibitemShut
  {NoStop}%
\bibitem [{\citenamefont {Lousto}\ and\ \citenamefont
  {Healy}(2019)}]{Lousto:2018dgd}%
  \BibitemOpen
  \bibfield  {author} {\bibinfo {author} {\bibfnamefont {C.~O.}\ \bibnamefont
  {Lousto}}\ and\ \bibinfo {author} {\bibfnamefont {J.}~\bibnamefont {Healy}},\
  }\href {\doibase 10.1103/PhysRevD.99.064023} {\bibfield  {journal} {\bibinfo
  {journal} {Phys. Rev.}\ }\textbf {\bibinfo {volume} {D99}},\ \bibinfo {pages}
  {064023} (\bibinfo {year} {2019})},\ \Eprint
  {http://arxiv.org/abs/1805.08127} {arXiv:1805.08127 [gr-qc]} \BibitemShut
  {NoStop}%
\bibitem [{\citenamefont {Healy}\ \emph {et~al.}(2014)\citenamefont {Healy},
  \citenamefont {Lousto},\ and\ \citenamefont {Zlochower}}]{Healy:2014yta}%
  \BibitemOpen
  \bibfield  {author} {\bibinfo {author} {\bibfnamefont {J.}~\bibnamefont
  {Healy}}, \bibinfo {author} {\bibfnamefont {C.~O.}\ \bibnamefont {Lousto}}, \
  and\ \bibinfo {author} {\bibfnamefont {Y.}~\bibnamefont {Zlochower}},\ }\href
  {\doibase 10.1103/PhysRevD.90.104004} {\bibfield  {journal} {\bibinfo
  {journal} {Phys. Rev.}\ }\textbf {\bibinfo {volume} {D90}},\ \bibinfo {pages}
  {104004} (\bibinfo {year} {2014})},\ \Eprint {http://arxiv.org/abs/1406.7295}
  {arXiv:1406.7295 [gr-qc]} \BibitemShut {NoStop}%
\bibitem [{\citenamefont {Healy}\ and\ \citenamefont
  {Lousto}(2017)}]{Healy:2016lce}%
  \BibitemOpen
  \bibfield  {author} {\bibinfo {author} {\bibfnamefont {J.}~\bibnamefont
  {Healy}}\ and\ \bibinfo {author} {\bibfnamefont {C.~O.}\ \bibnamefont
  {Lousto}},\ }\href {\doibase 10.1103/PhysRevD.95.024037} {\bibfield
  {journal} {\bibinfo  {journal} {Phys. Rev.}\ }\textbf {\bibinfo {volume}
  {D95}},\ \bibinfo {pages} {024037} (\bibinfo {year} {2017})},\ \Eprint
  {http://arxiv.org/abs/1610.09713} {arXiv:1610.09713 [gr-qc]} \BibitemShut
  {NoStop}%
\bibitem [{\citenamefont {Healy}\ \emph
  {et~al.}(2017{\natexlab{a}})\citenamefont {Healy}, \citenamefont {Lousto},\
  and\ \citenamefont {Zlochower}}]{Healy:2017mvh}%
  \BibitemOpen
  \bibfield  {author} {\bibinfo {author} {\bibfnamefont {J.}~\bibnamefont
  {Healy}}, \bibinfo {author} {\bibfnamefont {C.~O.}\ \bibnamefont {Lousto}}, \
  and\ \bibinfo {author} {\bibfnamefont {Y.}~\bibnamefont {Zlochower}},\ }\href
  {\doibase 10.1103/PhysRevD.96.024031} {\bibfield  {journal} {\bibinfo
  {journal} {Phys. Rev.}\ }\textbf {\bibinfo {volume} {D96}},\ \bibinfo {pages}
  {024031} (\bibinfo {year} {2017}{\natexlab{a}})},\ \Eprint
  {http://arxiv.org/abs/1705.07034} {arXiv:1705.07034 [gr-qc]} \BibitemShut
  {NoStop}%
\bibitem [{\citenamefont {Healy}\ \emph {et~al.}(2018)\citenamefont {Healy}
  \emph {et~al.}}]{Healy:2017abq}%
  \BibitemOpen
  \bibfield  {author} {\bibinfo {author} {\bibfnamefont {J.}~\bibnamefont
  {Healy}} \emph {et~al.},\ }\href {\doibase 10.1103/PhysRevD.97.064027}
  {\bibfield  {journal} {\bibinfo  {journal} {Phys. Rev.}\ }\textbf {\bibinfo
  {volume} {D97}},\ \bibinfo {pages} {064027} (\bibinfo {year} {2018})},\
  \Eprint {http://arxiv.org/abs/1712.05836} {arXiv:1712.05836 [gr-qc]}
  \BibitemShut {NoStop}%
\bibitem [{\citenamefont {Rosato}\ \emph {et~al.}(2021)\citenamefont {Rosato},
  \citenamefont {Healy},\ and\ \citenamefont {Lousto}}]{Rosato:2021jsq}%
  \BibitemOpen
  \bibfield  {author} {\bibinfo {author} {\bibfnamefont {N.}~\bibnamefont
  {Rosato}}, \bibinfo {author} {\bibfnamefont {J.}~\bibnamefont {Healy}}, \
  and\ \bibinfo {author} {\bibfnamefont {C.~O.}\ \bibnamefont {Lousto}},\
  }\href {\doibase 10.1103/PhysRevD.103.104068} {\bibfield  {journal} {\bibinfo
   {journal} {Phys. Rev. D}\ }\textbf {\bibinfo {volume} {103}},\ \bibinfo
  {pages} {104068} (\bibinfo {year} {2021})},\ \Eprint
  {http://arxiv.org/abs/2103.09326} {arXiv:2103.09326 [gr-qc]} \BibitemShut
  {NoStop}%
\bibitem [{\citenamefont {Zlochower}\ \emph {et~al.}(2012)\citenamefont
  {Zlochower}, \citenamefont {Ponce},\ and\ \citenamefont
  {Lousto}}]{Zlochower:2012fk}%
  \BibitemOpen
  \bibfield  {author} {\bibinfo {author} {\bibfnamefont {Y.}~\bibnamefont
  {Zlochower}}, \bibinfo {author} {\bibfnamefont {M.}~\bibnamefont {Ponce}}, \
  and\ \bibinfo {author} {\bibfnamefont {C.~O.}\ \bibnamefont {Lousto}},\
  }\href {\doibase 10.1103/PhysRevD.86.104056} {\bibfield  {journal} {\bibinfo
  {journal} {Phys. Rev.}\ }\textbf {\bibinfo {volume} {D86}},\ \bibinfo {pages}
  {104056} (\bibinfo {year} {2012})},\ \Eprint {http://arxiv.org/abs/1208.5494}
  {arXiv:1208.5494 [gr-qc]} \BibitemShut {NoStop}%
\bibitem [{\citenamefont {Healy}\ \emph
  {et~al.}(2017{\natexlab{b}})\citenamefont {Healy}, \citenamefont {Lousto},
  \citenamefont {Zlochower},\ and\ \citenamefont {Campanelli}}]{Healy:2017psd}%
  \BibitemOpen
  \bibfield  {author} {\bibinfo {author} {\bibfnamefont {J.}~\bibnamefont
  {Healy}}, \bibinfo {author} {\bibfnamefont {C.~O.}\ \bibnamefont {Lousto}},
  \bibinfo {author} {\bibfnamefont {Y.}~\bibnamefont {Zlochower}}, \ and\
  \bibinfo {author} {\bibfnamefont {M.}~\bibnamefont {Campanelli}},\ }\href
  {\doibase 10.1088/1361-6382/aa91b1} {\bibfield  {journal} {\bibinfo
  {journal} {Class. Quant. Grav.}\ }\textbf {\bibinfo {volume} {34}},\ \bibinfo
  {pages} {224001} (\bibinfo {year} {2017}{\natexlab{b}})},\ \Eprint
  {http://arxiv.org/abs/1703.03423} {arXiv:1703.03423 [gr-qc]} \BibitemShut
  {NoStop}%
\bibitem [{\citenamefont {Healy}\ \emph {et~al.}(2019)\citenamefont {Healy},
  \citenamefont {Lousto}, \citenamefont {Lange}, \citenamefont {O'Shaughnessy},
  \citenamefont {Zlochower},\ and\ \citenamefont {Campanelli}}]{Healy:2019jyf}%
  \BibitemOpen
  \bibfield  {author} {\bibinfo {author} {\bibfnamefont {J.}~\bibnamefont
  {Healy}}, \bibinfo {author} {\bibfnamefont {C.~O.}\ \bibnamefont {Lousto}},
  \bibinfo {author} {\bibfnamefont {J.}~\bibnamefont {Lange}}, \bibinfo
  {author} {\bibfnamefont {R.}~\bibnamefont {O'Shaughnessy}}, \bibinfo {author}
  {\bibfnamefont {Y.}~\bibnamefont {Zlochower}}, \ and\ \bibinfo {author}
  {\bibfnamefont {M.}~\bibnamefont {Campanelli}},\ }\href {\doibase
  10.1103/PhysRevD.100.024021} {\bibfield  {journal} {\bibinfo  {journal}
  {Phys. Rev.}\ }\textbf {\bibinfo {volume} {D100}},\ \bibinfo {pages} {024021}
  (\bibinfo {year} {2019})},\ \Eprint {http://arxiv.org/abs/1901.02553}
  {arXiv:1901.02553 [gr-qc]} \BibitemShut {NoStop}%
\bibitem [{\citenamefont {Healy}\ and\ \citenamefont
  {Lousto}(2022)}]{Healy:2022wdn}%
  \BibitemOpen
  \bibfield  {author} {\bibinfo {author} {\bibfnamefont {J.}~\bibnamefont
  {Healy}}\ and\ \bibinfo {author} {\bibfnamefont {C.~O.}\ \bibnamefont
  {Lousto}},\ }\href {\doibase 10.1103/PhysRevD.105.124010} {\bibfield
  {journal} {\bibinfo  {journal} {Phys. Rev. D}\ }\textbf {\bibinfo {volume}
  {105}},\ \bibinfo {pages} {124010} (\bibinfo {year} {2022})},\ \Eprint
  {http://arxiv.org/abs/2202.00018} {arXiv:2202.00018 [gr-qc]} \BibitemShut
  {NoStop}%
\bibitem [{\citenamefont {{B}rill}\ and\ \citenamefont
  {Lindquist}(1963)}]{Brill63}%
  \BibitemOpen
  \bibfield  {author} {\bibinfo {author} {\bibfnamefont {D.}~\bibnamefont
  {{B}rill}}\ and\ \bibinfo {author} {\bibfnamefont {R.}~\bibnamefont
  {Lindquist}},\ }\href@noop {} {\bibfield  {journal} {\bibinfo  {journal}
  {Phys. Rev.}\ }\textbf {\bibinfo {volume} {131}},\ \bibinfo {pages} {471}
  (\bibinfo {year} {1963})}\BibitemShut {NoStop}%
\bibitem [{\citenamefont {Lousto}\ and\ \citenamefont
  {Price}(1997{\natexlab{a}})}]{Lousto:1996sx}%
  \BibitemOpen
  \bibfield  {author} {\bibinfo {author} {\bibfnamefont {C.~O.}\ \bibnamefont
  {Lousto}}\ and\ \bibinfo {author} {\bibfnamefont {R.~H.}\ \bibnamefont
  {Price}},\ }\href@noop {} {\bibfield  {journal} {\bibinfo  {journal} {Phys.
  Rev. D}\ }\textbf {\bibinfo {volume} {55}},\ \bibinfo {pages} {2124}
  (\bibinfo {year} {1997}{\natexlab{a}})},\ \Eprint
  {http://arxiv.org/abs/gr-qc/9609012} {gr-qc/9609012} \BibitemShut {NoStop}%
\bibitem [{\citenamefont {Lousto}\ and\ \citenamefont
  {Zlochower}(2013)}]{Lousto:2013oza}%
  \BibitemOpen
  \bibfield  {author} {\bibinfo {author} {\bibfnamefont {C.~O.}\ \bibnamefont
  {Lousto}}\ and\ \bibinfo {author} {\bibfnamefont {Y.}~\bibnamefont
  {Zlochower}},\ }\href {\doibase 10.1103/PhysRevD.88.024001} {\bibfield
  {journal} {\bibinfo  {journal} {Phys. Rev.}\ }\textbf {\bibinfo {volume}
  {D88}},\ \bibinfo {pages} {024001} (\bibinfo {year} {2013})},\ \Eprint
  {http://arxiv.org/abs/1304.3937} {arXiv:1304.3937 [gr-qc]} \BibitemShut
  {NoStop}%
\bibitem [{\citenamefont {Campanelli}\ and\ \citenamefont
  {Lousto}(1999)}]{Campanelli:1998jv}%
  \BibitemOpen
  \bibfield  {author} {\bibinfo {author} {\bibfnamefont {M.}~\bibnamefont
  {Campanelli}}\ and\ \bibinfo {author} {\bibfnamefont {C.~O.}\ \bibnamefont
  {Lousto}},\ }\href {\doibase 10.1103/PhysRevD.59.124022} {\bibfield
  {journal} {\bibinfo  {journal} {Phys. Rev.}\ }\textbf {\bibinfo {volume}
  {D59}},\ \bibinfo {pages} {124022} (\bibinfo {year} {1999})},\ \Eprint
  {http://arxiv.org/abs/gr-qc/9811019} {arXiv:gr-qc/9811019 [gr-qc]}
  \BibitemShut {NoStop}%
\bibitem [{\citenamefont {Lousto}\ and\ \citenamefont
  {Price}(1997{\natexlab{b}})}]{Lousto:1997wf}%
  \BibitemOpen
  \bibfield  {author} {\bibinfo {author} {\bibfnamefont {C.~O.}\ \bibnamefont
  {Lousto}}\ and\ \bibinfo {author} {\bibfnamefont {R.~H.}\ \bibnamefont
  {Price}},\ }\href@noop {} {\bibfield  {journal} {\bibinfo  {journal} {Phys.
  Rev. D}\ }\textbf {\bibinfo {volume} {56}},\ \bibinfo {pages} {6439}
  (\bibinfo {year} {1997}{\natexlab{b}})},\ \Eprint
  {http://arxiv.org/abs/gr-qc/9705071} {gr-qc/9705071} \BibitemShut {NoStop}%
\bibitem [{\citenamefont {Lousto}\ and\ \citenamefont
  {Price}(2004)}]{Lousto:2004pr}%
  \BibitemOpen
  \bibfield  {author} {\bibinfo {author} {\bibfnamefont {C.~O.}\ \bibnamefont
  {Lousto}}\ and\ \bibinfo {author} {\bibfnamefont {R.~H.}\ \bibnamefont
  {Price}},\ }\href@noop {} {\bibfield  {journal} {\bibinfo  {journal} {Phys.
  Rev.}\ }\textbf {\bibinfo {volume} {D69}},\ \bibinfo {pages} {087503}
  (\bibinfo {year} {2004})},\ \Eprint {http://arxiv.org/abs/gr-qc/0401045}
  {gr-qc/0401045} \BibitemShut {NoStop}%
\bibitem [{\citenamefont {Lousto}(2005)}]{Lousto:2005ip}%
  \BibitemOpen
  \bibfield  {author} {\bibinfo {author} {\bibfnamefont {C.~O.}\ \bibnamefont
  {Lousto}},\ }\href@noop {} {\bibfield  {journal} {\bibinfo  {journal} {Class.
  Quant. Grav.}\ }\textbf {\bibinfo {volume} {22}},\ \bibinfo {pages} {S543}
  (\bibinfo {year} {2005})},\ \Eprint {http://arxiv.org/abs/gr-qc/0503001}
  {gr-qc/0503001} \BibitemShut {NoStop}%
\bibitem [{\citenamefont {Healy}\ \emph
  {et~al.}(2020{\natexlab{b}})\citenamefont {Healy}, \citenamefont {Lousto},\
  and\ \citenamefont {Rosato}}]{Healy:2020iuc}%
  \BibitemOpen
  \bibfield  {author} {\bibinfo {author} {\bibfnamefont {J.}~\bibnamefont
  {Healy}}, \bibinfo {author} {\bibfnamefont {C.~O.}\ \bibnamefont {Lousto}}, \
  and\ \bibinfo {author} {\bibfnamefont {N.}~\bibnamefont {Rosato}},\ }\href
  {\doibase 10.1103/PhysRevD.102.024040} {\bibfield  {journal} {\bibinfo
  {journal} {Phys. Rev. D}\ }\textbf {\bibinfo {volume} {102}},\ \bibinfo
  {pages} {024040} (\bibinfo {year} {2020}{\natexlab{b}})},\ \Eprint
  {http://arxiv.org/abs/2003.02286} {arXiv:2003.02286 [gr-qc]} \BibitemShut
  {NoStop}%
\end{thebibliography}%


%

\end{document}